\documentclass[twocolumn,showpacs,preprintnumbers,amsmath,amssymb]{revtex4}

\usepackage{graphicx}
\usepackage{dcolumn}
\usepackage{bm}

\begin{document}


\title{Self-similar solution of a nonsteady problem \\
of nonisothermal vapour condensation on a droplet \\
growing in diffusion regime}

\author{A. P. Grinin}
\author{G. Yu. Gor}
 \email{gennady_gor@mail.ru}
\author{F. M. Kuni}
\affiliation{
Saint-Petersburg state university, Research institute of physics \\
198504 Russia, St. Petersburg, Petrodvorets, Ulyanovskaya st., 1
}%

\date{\today}

\begin{abstract}
This paper presents a mathematically exact self-similar solution to the joint nonsteady problems of vapour diffusion towards a droplet growing in a vapour-gas medium and of removal of heat released by a droplet into a vapour-gas medium during vapour condensation. An equation for the temperature of the droplet is obtained; and it is only at that temperature that the self-similar solution exists. This equation requires the constancy of the droplet temperature and even defines it unambiguously throughout the whole period of the droplet growth. In the case of strong display of heat effects, when the droplet growth rate decreases significantly, the equation for the temperature of the droplet is solved analytically. It is shown that the obtained temperature fully coincides with the one that settles in the droplet simultaneously with the settlement of its diffusion regime of growth. At the obtained temperature of the droplet the interrelated nonsteady vapour concentration and temperature profiles of the vapour-gas medium around the droplet are expressed in terms of initial (prior to the nucleation of the droplet) parameters of the vapour-gas medium. The same parameters are used to formulate the law in accordance with which the droplet is growing in diffusion regime, and also to define the time that passes after the nucleation of the droplet till the settlement of diffusion regime of droplet growth, when the squared radius of the droplet becomes proportionate to time. For the sake of completeness the case of weak display of heat effects is been studied.
\end{abstract}

\maketitle

\section{Introduction}
\label{Introduction}

The problem of growth via diffusion of a solitary droplet in a supersaturated vapour was a subject of numerous papers 
 \cite{Reiss_LaMer,Reiss1951,Frisch1,Frisch2,Fuchs,Goodrich,Fuchs_Sutugin,Nix_Fukuta,Reiss1977,Grinina,Vasilyev}. 
This problem is an important constituent of the whole theory of phase transitions kinetics
 \cite{Tunitskii,Todes,Wakeshima,Kuni_Grinin,Slezov}. 
The solution of this problem is crucial for a thorough understanding of the ensemble evolution of droplets competing for consumption of supersaturated vapour.

When considering a solitary droplet, primary issues under study are the time dependence of the droplet radius and the concentration profile around the droplet. As far back as in the paper by Reiss and La Mer \cite{Reiss_LaMer} it was noted that the main mathematical difficulty in the description of droplet growth is related with the boundary condition on its moving surface. Traditionally (e.g. see \cite{Fuchs}) the problem of vapour diffusion towards a growing droplet was treated in two steps. On the first step one obtained vapour concentration profile around the droplet with a fixed radius and consequently  the diffusion flux. On the second step, making use of the diffusion flux value determined on the first step, the growth rate of a droplet radius was determined; and therefore  the radius as a function of time. When a fixed radius in the expression for the vapour concentration profile was replaced with a time-dependant radius obtained on the second step, the result was considered as an approximation for the exact solution of the diffusion problem. The shortcoming of the conventional solution is that the balance of the condensing matter is not maintained. This shortcoming was demonstrated recently in the paper \cite{Grinina}. \cite{Grinina} presented an approximate solution of the diffusion problem complying with the balance of the condensing matter. 

In the paper by Vasilyev et al. \cite{Vasilyev} another treatment of the diffusion problem was realized. The observance of the balance of the condensing matter was accepted a priori. Then, using dimensional arguments, it was shown that the problem could be reformulated in terms of one dimensionless variable (instead of two variables -- spherical coordinate and time). Thus in \cite{Vasilyev} the mathematically exact self-similar solution of the problem of vapour diffusion to the growing droplet was obtained.

To be fair it has to be noted that the reduction of a spherically symmetric problem of diffusion of matter towards a growing particle of a new phase to one variable was performed earlier independently by Zener \cite{Zener} and Frank \cite{Frank}. In both \cite{Zener} and \cite{Frank} the process under consideration was the growth of a crystal in a solution. Fuchs \cite{Fuchs} regarded Frank's results as incorrect; and probably that was the reason these papers did not receive recognition in USSR.

Despite the fact that the results from papers \cite{Grinina, Vasilyev} are highly accurate from a mathematical point of view, there are very few possibilities to use them for practical purposes. The problem is that, during the condensational growth of the droplet, latent heat of condensation is released, which causes heating of the droplet and the medium around it. For these effects to be neglected, as it was done in \cite{Grinina, Vasilyev}, it is required that a significant amount of an inert gas functioning as a thermostat is present. This condition is difficult to satisfy in practice.

The main aim of the present paper is to take into account effects of condensation heat release during the diffusion growth of a droplet in a vapour-gas medium. Here, besides the problem of vapour diffusion towards a growing droplet, a related heat conductivity problem emerges. The finding of solutions of joint diffusion and heat conductivity problems around the droplet growing in a supersaturated vapour were presented in numerous papers. In \cite{Kuni1985} these problems were considered in a steady approximation. In \cite{Fuchs, Pesthy} nonsteady solutions were presented; however, these solutions had an essential shortcoming which was further treated in \cite{Grinina}. In \cite{Nix_Fukuta} joint diffusion and heat conductivity problems were considered with specific time-dependent conditions of heat and matter fluxes into the system. In \cite{Frank} nonsteady diffusion and heat conductivity problems were solved for the case of crystal growth. Frank \cite{Frank} exploited the fact that the coefficient of thermal diffusivity in the liquid medium significantly exceeds diffusion coefficient. This is not true for the case of a vapour-gas medium.

The present paper is devoted to the determination of a self-similar solution of joint nonsteady problems of vapour diffusion towards a droplet growing nonisothermally in a vapour-gas medium and of abstraction of heat released by the droplet during vapour condensation into the vapour-gas medium.

The basis for this will be the balance equation on the number of molecules released by diffusion from vapour by the current time and the number of molecules that have composed a growing droplet. Although the observance of the balance equation will be required at all times after the nucleation of the droplet, the self-similar solution obtained on its basis will only refer to times when the droplet is growing in diffusion regime.

An equation for the temperature of the droplet will be obtained; and it is only at that temperature that the self-similar solution exists. This equation will require the constancy of the droplet temperature and even will define it unambiguously throughout the whole period of the droplet growth.

In case of strong display of heat effects, when the heating of a droplet by condensation heat causes a significant decrease in its growth rate, the equation for the temperature of the droplet will be solved analytically. It will be shown that the obtained temperature fully coincides with the one that, in accordance with \cite{Kuni1985}, settles in the droplet simultaneously with the settlement of diffusion regime of its growth. Constant temperature of the growing droplet, which is required by self-similar theory, is consequently prominent from a physical standpoint, or stable, to be more exact.

At the obtained temperature of the droplet the interrelated nonsteady vapour concentration and temperature profiles of the vapour-gas medium around the droplet, those that have not been studied earlier, will be expressed in terms of initial (prior to the droplet nucleation) parameters of the vapour-gas medium. The same parameters will be used to express the law in accordance with which the droplet is growing in diffusion regime, and also to express the previously undiscussed time that passes after the nucleation of the droplet till the settlement of diffusion regime of the droplet growth, when the squared radius of the droplet becomes proportionate to time.

Strong display of heat effects is possible only when the initial temperature of vapour differs considerably from the critical point liquid-vapour. It is in this case that the dimensionless condensation heat (referred to one molecule and expressed in heat units of energy) significantly exceeds one. In the present paper it will be shown that strong display of heat effects requires only that the squared dimensionless condensation heat, which exceeds the heat itself significantly, would also exceed the ratio between the quantity of the inert gas in the vapour-gas medium and the initial quantity of vapour. In this case the limitation from above set on the ratio between the quantities of the inert gas and vapour will be rather weak. Consequently, this bound is compatible with the requirement needed for the description of diffusion and heat conductivity used in the present paper, i.e. the ratio between the quantities of the inert gas and vapour would significantly exceed one.

For the sake of completeness this paper will also examine a case when the ratio between the quantities of the inert gas and vapour is rather large and heat effects are displayed weakly.

We should emphasize that the impetus for the creation of papers \cite{Grinina, Vasilyev} was given by the elaboration of the so-called ``statistico-probabilistic approach'' to taking account of vapour depletion in the kinetics of homogeneous nucleation for the free-molecular regime
of droplet growth \cite{Djikaev} and its further generalization for the case of diffusion regime \cite{Reiss}. The essence of this approach, also denoted in \cite{Reiss} as a ``nearest-neighbour approximation'', is the consideration of the influence of a growing droplet on the nucleation of the droplet nearest to it. To realize this approach one needs to find the concentration profile around the growing droplet with high accuracy.

An important notion of the conventional scheme of the description of nucleation
 \cite{Tunitskii,Todes,Wakeshima,Kuni_Grinin,Slezov}
is the concept of collective consumption of excess vapour by growing droplets. That concept implies the high degree of spatial uniformity of the system. In fact every droplet does not consume vapour uniformly from the whole system volume, but it does so mainly from its vicinity (named in \cite{Pesthy} as ``clearance volume''). The probability of nucleation of the next droplet in the vicinity of the first one is negligible. In papers \cite{Djikaev, Reiss} the process of nucleation of a droplet which is the nearest to the given growing droplet is investigated. Under certain conditions the mean distance to the nearest-neighbour drop and the mean time to its appearance can be determined reliably. Under these conditions, the mean time provides an estimate for the duration of the nucleation stage, while the mean distance provides an estimate for the number of drops that formed per unit volume during the nucleation stage.

The solution of joint nonsteady diffusion and heat conductivity problems suggested in the present paper can serve as a step on the way to the generalization of the ``nearest-neighbour approximation'' \cite{Djikaev, Reiss} to the nonisothermal case.

\section{Nonsteady vapour concentration profile around the droplet growing in diffusion regime}
\label{Concentration}

Let us consider initially homogeneous vapour-gas mixture at absolute temperature $T_0$ which contains $n_0$ molecules of vapour in one unit of volume. Vapour is supersaturated, so $n_0$ exceeds the concentration (number density of molecules) of saturated vapour at the plain surface of the liquid phase $n_{\infty}(T_0)$ at temperature $T_0$: $n_0 > n_{\infty}(T_0)$. A supercritical droplet nucleates and begins growing in the vapour, condensing the surrounding vapour particles and simultaneously releasing condensation heat. Let us consider such conditions when the droplet reaches the sizes significantly exceeding the free path length $\lambda $ of a vapour molecule in the vapour-gas medium. Under these conditions the transfer of vapour molecules is governed by the diffusion equation, while heat transfer in the vapour-gas medium -- by the equation of heat conductivity. It should be noted that in this paper, unlike in \cite{Kuni1985}, we are going to consider nonsteady transport phenomena.

Let us assume that all the heat released on a droplet in the process of its growth will be completely drawn-off to the vapour-gas medium. This assumption will be substantiated for in \cite{Kuni1985}. The moment of the nucleation of the droplet will be taken as a time reference point $t = 0$. We will designate the radius of the droplet growing with time as $R(t)$.

Let $R_{D} $ be such droplet radius at which the diffusion regime of droplet growth with proportional dependence of the squared radius of the droplet on time can be considered to be settled. Radius $R_{D}$ will be determined by the equation
\begin{equation}
\label{1.1}
R_D = N \lambda.
\end{equation}
There is arbitrary rule in the definition of multiplier $N ~(N \gg 1)$; in our estimates we will put it that $N \sim 10$. Let us introduce characteristic time $t_{D}$ at which the radius of the droplet satisfies the initial condition
\begin{equation}
\label{1.2}
R \left(t \right) \big |_{t = t_{D}} =R_{D}.
\end{equation}
In accordance with Eq. \eqref{1.2} time $t_{D}$ can be interpreted as time of the settlement of diffusion regime of the droplet growth counted from the moment of the nucleation of the droplet, at which the squared radius of the droplet becomes proportional to time. Further, when solving the transport problem, we will be dealing with times $t \ge t_{D} $.

The description of diffusion and heat conductivity used further in the paper supposes that the concentration of the inert gas $n_{g}$ in the vapour-gas mixture significantly exceeds the initial concentration of vapour $n_0$:
\begin{equation}
\label{1.3}
n_{g} \gg n_{0}.
\end{equation}
Inequality Eq. \eqref{1.3}, however, should not be too strong. Otherwise, the effect of condensation heat release we are interested in will not reveal itself. Section \ref{Strong} will include explicit formulation of the upper bound on the ratio $n_g/n_0$, at which the effect is going to be strong. Inequality Eq. \eqref{1.3} makes it possible to neglect the difference between the free path length of a vapour molecule in the vapour-gas medium and in the inert gas.

The concentration of vapour $n(r,t)$ at distance $r$ from the centre of the droplet at time $t$ can be obtained by solving the diffusion equation.
\begin{equation}
\label{1.4}
\frac{\partial n\left(r,t\right)}{\partial t} =\frac{D}{r^{2} } \frac{\partial }{\partial r} \left(r^{2} \frac{\partial n\left(r,t\right)}{\partial r} \right),
\end{equation}
where $D$ is the coefficient of diffusion of vapour molecules in the inert gas. Coefficient $D$ depends on the concentration of the inert gas $n_g$, as well as free path length $\lambda$, in inverse proportion to $n_g$. Eq. \eqref{1.4} is provided with the initial condition of homogeneity of vapour concentration profile
\begin{equation}
\label{1.5}
n \left(r,t\right) \big|_{t=0} = n_{0},
\end{equation}
equilibrium boundary condition on the surface of the droplet
\begin{equation}
\label{1.6}
n \left(r,t\right) \big |_{r=R\left(t\right)} = n_{\infty } \left(T_{d} \right)
\end{equation}
and a boundary condition of the absence of an external vapour source
\begin{equation}
\label{1.7}
\left[r^{2} \frac{\partial n\left(r,t\right)}{\partial r} \right] \bigg|_{r=\infty } = 0,
\end{equation}
where $n_{\infty } \left(T_{d} \right)$ is the concentration of saturated vapour at the plain surface of the liquid in the droplet at the temperature of the droplet $T_{d}$. The surface can be considered plain as the radius of the droplet $R$ exceeds free path length $\lambda$ significantly. Due to high density of liquid in the droplet in comparison to the density of vapour, value $n_{\infty } \left(T_{d} \right)$, as well as value $n_{\infty } \left(T_{0} \right)$, is hyposensitive to the pressure that the inert gas exerts on the liquid in the droplet.

We will further assume that the temperature of the growing droplet $T_{d} $ does no depend on its size and, consequently, on time. This assumption will be confirmed further. Moreover, we will obtain an equation that will unambiguously determine the temperature of the droplet $T_{d} $, which is constant in time. The settlement of the constant in time temperature of the growing droplet is beyond the scope of the self-similar theory. Earlier this settlement was studied in \cite{Kuni1985}, where it was shown that it takes place simultaneously with the settlement of the diffusion regime of the droplet growth.

The important role will be played by the equation on the balance of number of molecules of the condensing substance:
\begin{equation}
\label{1.8}
\frac{4\pi }{3} R^{3} \left(t\right)\left[n_{l} -n_{0} \right]=4\pi \int _{R\left(t\right)}^{\infty }drr^{2}  \left[n_{0} -n\left(r,t\right)\right],
\end{equation}
where $n_{l} $ is the number density of molecules in the droplet.

Let us validate this equation. Let us imagine a spherical surface of radius $R\left(t\right)$ around the centre of the droplet. At the reference time $t=0$, in concord with the initial condition Eq. \eqref{1.5}, inside the surface of radius $R\left(t\right)$ there was a homogeneous vapour-gas mixture with vapour concentration $n_{0} $. At the time moment $t\ge t_{D}$ inside the same surface there is a droplet with the concentration $n_{l} $. The difference between the number of molecules of the condensing substance inside the spherical surface of radius $R\left(t\right)$ at the time $t\ge t_{D}$ and the corresponding number of molecules at the time $t=0$ is given by the left side of the equation Eq. \eqref{1.8}. This difference corresponds to the influx of molecules of the condensing substance inside the spherical surface of radius $R\left(t\right)$.

On the other hand, the difference between the number of molecules of the condensing substance outside the spherical surface of radius $R\left(t\right)$ at the time $t=0$ and the corresponding number of molecules at the time $t\ge t_{D} $ is given by the right side of the equation Eq. \eqref{1.8}. This difference corresponds to the escape of molecules from vapor-gas mixture outside the spherical surface of radius $R\left(t\right)$. In conditions of physical isolation of the system, which is expressed by a boundary condition Eq. \eqref{1.7}, this difference also corresponds to the influx of vapour molecules inside the spherical surface of radius $R\left(t\right)$, i.e. is equal to the left side of the equation Eq. \eqref{1.8}. This very fact becomes the substantiation of the balance equation Eq. \eqref{1.8}.

Differentiating equation Eq. \eqref{1.8} with respect to $t$ at $t\ge t_{D}$, we have
\begin{eqnarray}
\label{1.9}
R^{2} \left(t\right)\frac{dR\left(t\right)}{dt} \left[n_{l} -n_{0} \right]=-R^{2} \left(t\right)\frac{dR\left(t\right)}{dt} \nonumber \\
\times \left[n_{0} -\left. n\left(r,t\right)\right|_{r=R\left(t\right)} \right] 
-\int _{R\left(t\right)}^{\infty }drr^{2}  \frac{\partial n\left(r,t\right)}{\partial t}.
\end{eqnarray}

Using the boundary condition Eq. \eqref{1.6} and engaging Eq. \eqref{1.4} when calculating the derivative of $n(r,t)$ on the right side Eq. of \eqref{1.9}, we will rewrite Eq. \eqref{1.9} as
\begin{equation}
\label{1.10}
R^{2} \left(t\right)\frac{dR\left(t\right)}{dt} \left[n_{l} -n_{\infty } \left(T_{d} \right)\right]=-D\, \left. \left[r^{2} \frac{\partial n\left(r,t\right)}{\partial r} \right] \right|_{R\left(t\right)}^{\infty }.
\end{equation}
If, when dealing with Eq. \eqref{1.10}, we take into account the boundary condition Eq. \eqref{1.7}, we can derive an expression for the rate of the droplet radius change
\begin{equation}
\label{1.11}
\frac{dR\left(t\right)}{dt} =\frac{D}{n_{l} - n_{\infty } \left(T_{d} \right)} \left. \frac{\partial n\left(r,t\right)}{\partial r} \right|_{r=R\left(t\right)} .
\end{equation}

Following \cite{Vasilyev}, we will introduce a self-similar variable $\rho$ and define it as follows 
\begin{equation}
\label{1.12}
\rho =r/R(t) ~~~~~~~ (\rho \ge 1)
\end{equation}
We will seek a solution of the equation Eq. \eqref{1.4} in the form of a function $n(\rho)$ of one variable $\rho$:
\begin{equation}
\label{1.13}
n\left(r,t\right)=n\left(\rho \right).
\end{equation}
In accordance with Eqs. \eqref{1.12}, \eqref{1.13} for partial derivatives of $n(r, t)$ we have
\begin{eqnarray}
\label{1.14}
\frac{\partial n\left(r,t\right)}{\partial t} =-\frac{r}{R^{2} \left(t\right)} \frac{dR\left(t\right)}{dt} \frac{dn\left(\rho \right)}{d\rho } , \nonumber \\
\frac{\partial n\left(r,t\right)}{\partial r} =\frac{1}{R\left(t\right)} \frac{dn\left(\rho \right)}{d\rho }.
\end{eqnarray}
Using Eq. \eqref{1.11} and the second relation in Eq. \eqref{1.14}, we have
\begin{equation}
\label{1.15}
R\left(t\right)\frac{dR\left(t\right)}{dt} =\frac{D}{n_{l} -n_{\infty } \left(T_{d} \right)} \left. \frac{dn\left(\rho \right)}{d\rho } \right|_{\rho =1}.
\end{equation}

Let us rewrite Eq. \eqref{1.15} as
\begin{equation}
\label{1.16}
 R(t) \frac{dR(t)}{dt} = D b,
\end{equation}
or, equivalently, as
\begin{equation}
\label{1.17}
 dR^2(t)/dt = 2 D b,
\end{equation}
where a crucial dimensionless parameter $b$ is introduced by means of
\begin{equation}
\label{1.18}
 b \equiv \frac{1}{n_{l} - n_{\infty}(T_d)} \left. \frac{dn(\rho)}{d\rho} \right | _{\rho=1}.
\end{equation}
In accordance with Eq. \eqref{1.17} the derivative $dR^2(t)/dt$ does not depend on time: along the axis of variable $R^2$ the droplet is ``moving'' with a velocity that does not depend on time and on the sizes of the droplet itself.

Integrating Eq. \eqref{1.17} with the initial condition Eq. \eqref{1.2}, we have
\begin{equation}
\label{1.19}
R^2(t) = 2 D b t + (R_D^2 - 2 D b t_D) ~~~~~~~ (t \ge t_D).
\end{equation}
With the increase of time $t$ the role of the addend in parenthesis in Eq. \eqref{1.19} decreases, and the dependency $R^2(t)$ becomes compliant with the law
\begin{equation}
\label{1.20}
R^2(t) = 2 D b t ~~~~~~~ (t \ge t_D).
\end{equation}
Which allows us to write the following approximate equality
\begin{equation}
\label{1.21}
t_D \simeq \frac{R_D^2}{2 D b}.
\end{equation}
However, in order to estimate precisely the measure of inaccuracy of the approximate equality Eq. \eqref{1.21}, one still needs to know the law in accordance with which $R^2(t)$ changes with the increase of time within the interval $0 \le t \le t_D$. Let us take it into account that the growth of the droplet during the interval of $0 \le t \le t_D$ occurs not slower that in accordance with the law Eq. \eqref{1.20} (in free-molecular regime of growth the squared radius of the droplet is changing in proportion to the squared time, e.g. \cite{Djikaev}). Then we see that Eq. \eqref{1.21}, already when $N \sim 10$, allows us to find the approximate time $t_D$ which is needed to reach compliance with the law Eq. \eqref{1.20}, according to which the squared radius of the droplet grows in proportion to time.

It can be easily shown that, considering Eqs. \eqref{1.12} -- \eqref{1.14}, \eqref{1.16}, Eq. \eqref{1.4}, can be reduced to the form of an ordinary differential equation on the function $n(\rho)$ of the following type
\begin{equation}
\label{1.22}
\frac{d^{2} n\left(\rho \right)}{d\rho ^{2} } +\left(\frac{2}{\rho } +b\rho \right)\frac{dn\left(\rho \right)}{d\rho } =0.
\end{equation}
Eq. \eqref{1.22} can be easily integrated. An arbitrary variable that emerges after the first integration, considering Eq. \eqref{1.18}, can be expressed by means of parameter $b$ . An arbitrary variable that emerges after the second integration can be derived from the boundary condition Eq. \eqref{1.6}, which may be rewritten using the dimensionless variable $\rho$ in the following way:
\begin{equation}
\label{1.23}
\left. n\left(\rho \right)\right|_{\rho =1} =n_{\infty } \left(T_{d} \right).
\end{equation}
Thus equation Eq. \eqref{1.22} will have the following solution:
\begin{eqnarray}
\label{1.24}
n\left(\rho \right)=n_{\infty } \left(T_{d} \right)+b \left[ n_{l} -n_{\infty } \left(T_{d} \right)\right] \exp \left(\frac{b}{2} \right) \nonumber \\
\times \int _{1}^{\rho }\frac{dx}{x^{2} }  \exp \left(-\frac{b}{2} x^{2} \right).
\end{eqnarray}
Expression Eq. \eqref{1.24} for vapour concentration should refer only to times $t\ge t_{D} $. The initial condition Eq. \eqref{1.5} at the time $t=0$ that is important for the balance equation Eq. \eqref{1.8} thus loses its meaning in conformity with Eq. \eqref{1.24}. Despite this fact, it is evident from the initial condition Eq. \eqref{1.5} and from the boundary condition Eq. \eqref{1.7} that, if we take physical considerations into account, the following boundary condition will be justified: $\left. n(r,t) \right|_{r=\infty} = n_0$. In terms of variable $\rho$ it can be rewritten as follows
\begin{equation}
\label{1.25}
\left. n\left(\rho \right)\right|_{\rho =\infty } = n_{0}.
\end{equation}
Thus the initial concentration of vapour $n_0$ has the meaning of a constant in time concentration of vapour at the infinite separation from the droplet.

The boundary condition Eq. \eqref{1.25} provides the equation for the derivation of parameter $b$. Considering Eq. \eqref{1.24}, this equation looks as follows:
\begin{equation}
\label{1.26}
a =b \exp \left(\frac{b}{2} \right)\int _{1}^{\infty }\frac{dx}{x^{2} } \exp \left(-\frac{b}{2} x^{2} \right).
\end{equation}
Here an important dimensionless parameter $a$ is introduced. It is defined by an equality
\begin{equation}
\label{1.27}
a \equiv \frac{n_0 - n_{\infty}(T_d)}{n_l - n_{\infty}(T_d)}.
\end{equation}
The solution of the transcendental equation Eq. \eqref{1.26}, as may be shown, exists and is unique when the values of parameters comply with inequalities
\begin{equation}
\label{1.28}
0 < a < 1, ~~~~~ 0 < b < \infty.
\end{equation}
In order for Eq. \eqref{1.27} to enforce $a > 0$, the following inequality is needed: $n_0 - n_{\infty}(T_d) > 0$. The fact that this inequality follows from the inequality $n_0 > n_{\infty}(T_0)$ adopted in the beginning of the section will be justified further. In accordance with Eqs. \eqref{1.24} and \eqref{1.25} and \eqref{1.28}, with the increase of $\rho $ from 1 to $\infty$,  function $n(\rho)$ increases from $n_{\infty}(T_d)$ to $n_{0}$.

As can be seen from Eqs. \eqref{1.26}, \eqref{1.27}, in order to determine parameter $b$ one needs to know the temperature of the growing droplet $T_{d}$.

\section{Nonsteady temperature profile around the droplet growing in diffusion regime}
\label{Temperature}

Phase transition heat carried away from the droplet as the result of condensation of vapour molecules is drawn off to the vapour-gas medium. Temperature profile evolution $T\left(r,t\right)$ around the growing droplet complies with the equation of heat conductivity
\begin{equation}
\label{2.1}
\frac{\partial T\left(r,t\right)}{\partial t} = \frac{\chi}{r^{2} } \frac{\partial }{\partial r} \left(r^{2} \frac{\partial T\left(r,t\right)}{\partial r} \right)
\end{equation}
with the initial condition of temperature profile homogeneity
\begin{equation}
\label{2.2}
\left. T\left(r,t\right)\right|_{t=0} =T_{0} ,
\end{equation}
with the boundary condition of equality of medium temperature near the surface of the droplet and a currently unknown temperature of the droplet
\begin{equation}
\label{2.3}
\left. T\left(r,t\right)\right|_{r=R\left(t\right)} =T_{d}
\end{equation}
and with the boundary condition of the absence of an external heat source
\begin{equation}
\label{2.4}
\left. \left[r^{2} \frac{\partial T\left(r,t\right)}{\partial r} \right]\right|_{r=\infty } =0.
\end{equation}
Here, $\chi$ is the coefficient of thermal diffusivity of the inert gas. Strong inequality Eq. \eqref{1.3} allows us to neglect the differences between the coefficient of thermal diffusivity of the vapour-gas mixture and the coefficient of thermal diffusivity of the inert gas. Coefficient $\chi $, just as coefficient $D$, depends on the concentration of the inert gas $n_g$ in inverse proportion. Mathematical equivalence of the boundary value problem of heat conductivity, Eqs. \eqref{2.1} -- \eqref{2.4}, to the diffusion problem, Eqs. \eqref{1.4} --  \eqref{1.7}, allows it to be assumed that
\begin{equation}
\label{2.5}
T\left(r,t\right)=T\left(\rho \right),
\end{equation}
and, taking into account Eqs. \eqref{1.12} and \eqref{1.16}, it becomes possible at times $t \ge t_D$ to reduce the problem to the solution of an ordinary differential equation
\begin{equation}
\label{2.6}
\frac{d^{2} T\left(\rho \right)}{d\rho ^{2} } +\left(\frac{2}{\rho } +b \frac{D}{\chi} \rho \right)\frac{dT\left(\rho \right)}{d\rho } =0
\end{equation}
with boundary conditions
\begin{equation}
\label{2.7}
\left. T\left(\rho \right)\right|_{\rho =1} =T_{d} ,
\end{equation}
\begin{equation}
\label{2.8}
\left. T\left(\rho \right)\right|_{\rho =\infty } =T_{0} .
\end{equation}
The boundary condition Eq. \eqref{2.8}, analogous to the boundary condition Eq. \eqref{1.25}, follows, on the grounds of physics, from the initial condition Eq. \eqref{2.2} and from the boundary condition Eq. \eqref{2.4}. Thus, the initial temperature $T_0$ also carries the meaning of constant in time temperature at an infinite distance from the droplet.

Interconnection between the heat conductivity problem and the diffusion problem is realized by means of equality of heat released from the droplet during the process of vapour condensation and of heat removed from the droplet to the vapour-gas medium by means of heat conductivity. Let us put down this equality
\begin{equation}
\label{2.10}
q  j_{D} \left(t\right)=j_{q} \left(t\right).
\end{equation}
Here $q$ is the condensation heat per one molecule. The diffusion flux of vapour molecules towards the droplet $j_D$ and the heat flux $j_q$ removed from droplet into the medium are determined by the following equalities:
\begin{equation}
\label{2.11}
j_{D} \left(t\right)=4\pi r^{2} D\left. \frac{\partial n\left(r,t\right)}{\partial r} \right|_{r=R\left(t\right)} ,  \end{equation}
\begin{equation}
\label{2.12}
j_{q} \left(t\right)=-4\pi r^{2} \kappa \left. \frac{\partial T\left(r,t\right)}{\partial r} \right|_{r=R\left(t\right)} .  \end{equation}
The coefficient of heat conductivity of the inert gas $\kappa$ in Eq. \eqref{2.12} is related to the coefficient of thermal diffusivity of the inert gas $\chi$ by means of relation
\begin{equation}
\label{2.13}
\kappa =k_{B} n_{g} c_{g} \chi,
\end{equation}
where $k_B$ is the Boltzmann constant, $c_g$ is the molecular heat capacity of the inert gas at constant pressure measured in $k_B$ units. As $\chi$ relates to $n_{g}$ in inverse proportion, then the coefficient $\kappa$ does not depend on the concentration of the inert gas.

In Eqs. \eqref{2.11}, \eqref{2.12} let substitute derivatives by $r$ for derivatives by $\rho $. Using  Eq. \eqref{2.10},  we obtain the following relation
\begin{equation}
\label{2.14}
\left. \kappa \frac{dT\left(\rho \right)}{d\rho } \right|_{\rho =1} =-q D\left. \frac{dn\left(\rho \right)}{d\rho } \right|_{\rho =1} .
\end{equation}
Taking into account Eq. \eqref{1.18},  from Eq. \eqref{2.14} it follows
\begin{equation}
\label{2.14a}
\left. \frac{dT\left(\rho \right)}{d\rho } \right|_{\rho =1} =-\frac{q b D}{\kappa } \left[n_{l} -n_{\infty } \left(T_{d} \right)\right].
\end{equation}

The solution of Eq. \eqref{2.6} looks as follows:
\begin{eqnarray}
\label{2.15}
T\left(\rho \right)=T_{d} -\frac{qbD}{\kappa } \left[n_{l} -n_{\infty } \left(T_{d} \right)\right]\exp \left(\frac{b}{2} \frac{D}{\chi} \right) \nonumber \\ 
\times \int _{1}^{\rho }\frac{dx}{x^{2} }  \exp \left(-\frac{b}{2} \frac{D}{\chi} x^{2} \right).
\end{eqnarray}
In order to find the arbitrary constant that emerges as a result of the first integration, Eq. \eqref{2.14a} was made use of. The constant from the second integration was derived from the boundary condition Eq. \eqref{2.7}. Using the boundary condition Eq. \eqref{2.8}, from Eq. \eqref{2.15} we have
\begin{eqnarray}
\label{2.16}
T_{d} =T_{0} +\frac{qbD}{\kappa } \left[n_{l} -n_{\infty } \left(T_{d} \right)\right]\exp \left(\frac{b}{2} \frac{D}{\chi} \right) \nonumber \\
\times \int _{1}^{\infty }\frac{dx}{x^{2} }  \exp \left(-\frac{b}{2} \frac{D}{\chi} x^{2} \right).
\end{eqnarray}

Transcendental equation Eq. \eqref{2.16} unambiguously sets the temperature of the droplet $T_{d} $, which does not depend on the size of the droplet, and, consequently, on time. Eq. \eqref{2.16} is a condition that serves for the existence of self-similar solutions Eqs. \eqref{1.24} and \eqref{2.15}. From Eqs. \eqref{2.16} and \eqref{1.28} it can be seen that $T_{d} >T_{0} $. In accordance with Eqs. \eqref{2.15} and \eqref{2.16}, with the increase of $\rho $ from 1 to $\infty$, the temperature decreases from $T_d$ to $T_{0}$. When the flux of the substance in the vapour-gas medium is oriented towards the droplet, the heat flux, on the contrary, is headed away from the droplet. The fact that inequality $n_{0} -n_{\infty } \left(T_{d} \right)>0$, needed in accordance with Eqs. \eqref{1.27} and \eqref{1.28}, is observed simultaneously with the observance of $n_{0} -n_{\infty } \left(T_{0} \right)>0$, despite the inequality $n_{\infty } \left(T_{d} \right)>n_{\infty } \left(T_{0} \right)$, which follows from $T_{d} >T_{0} $, will become evident further from Eq. \eqref{4.5}.

Relations Eqs. \eqref{1.26} and \eqref{2.16} form a system of two transcendental equations that makes it possible to find the temperature of the droplet $T_{d} $ and parameter $b$ in the law Eq. \eqref{1.20} on the droplet growth .

\section{Approximation of high density of the liquid in the droplet}
\label{Density}

The system of Eqs. \eqref{1.26} and \eqref{2.16} can be solved numerically without any further stipulations. However, under the restriction that is fulfilled away from the vapour-gas critical point
\begin{equation}
\label{3.1}
n_{l} \gg n_{0}
\end{equation}
it is possible to obtain the solution of the system of equations Eqs. \eqref{1.26} and \eqref{2.16} analytically.

At the observance of a strong inequality Eq. \eqref{3.1}, from definition Eq. \eqref{1.27} it follows
\begin{equation}
\label{3.3}
a^{1/2} \ll 1.
\end{equation}
Differentiating by $b$, it is easy to make sure that that the following asymptotic formula is just
\begin{eqnarray}
\label{3.2}
\int _{1}^{\infty }\frac{dx}{x^{2} } \exp \left(-\frac{b}{2} x^{2} \right) =1-\pi ^{1/2}  \left(\frac{b}{2} \right)^{1/2} +\frac{b}{2} \nonumber \\
-\frac{1}{6} \left(\frac{b}{2} \right)^{2} +... ~~~ \left(\left({b/2} \right)^{1/2} \ll 1\right).
\end{eqnarray}
If we limit ourselves to the main order when expanding Eq. \eqref{3.2}, it is possible to show that Eq. \eqref{1.26} results in
\begin{equation}
\label{3.4}
b = a.
\end{equation}
The consequence of Eqs. \eqref{3.3} and \eqref{3.4} is the inequality $\left({b/2} \right)^{1/2} \ll 1$ presupposed in Eqs. \eqref{3.2} and \eqref{3.4}.
 
Considering that in case of gases the following estimate is justified
\begin{equation}
\label{3.5a}
D/\chi \sim 1,
\end{equation}
we can use Eq. \eqref{3.2}, having replaced $b$ with $bD/\chi$. From Eq. \eqref{2.16} we consequently have
\begin{equation}
\label{3.5b}
T_{d} =T_{0} +\frac{q b D}{\kappa } \left[n_{l} -n_{\infty } \left(T_{d} \right)\right].
\end{equation}
Taking Eqs. \eqref{3.4} and \eqref{1.27} into consideration, we can rewrite Eq. \eqref{3.5b} as
\begin{equation}
\label{3.5}
T_{d} =T_{0} +\frac{q D}{\kappa } \left[n_{0} -n_{\infty } \left(T_{d} \right)\right].
\end{equation}

Let us make sure that Eq. \eqref{3.5} on constant in time temperature of the droplet is precisely equivalent to the analogous equation obtained earlier in \cite{Kuni1985}. Making use of the Clapeyron-Clausius formula at $(T_{d} -T_{0})/T_{0} \ll 1 $, let us write
\begin{equation}
\label{3.6}
n_{\infty } \left(T_{d} \right)=n_{\infty } \left(T_{0} \right)\exp \left(\frac{q}{k_{B} T_{0} } \frac{T_{d} -T_{0} }{T_{0} } \right).
\end{equation}
In Eq. \eqref{3.6} it is supposed that vapour complies with the constitutive equation of ideal gas. Inequality $(T_{d} -T_{0})/T_{0} \ll 1 $ will be justified later.

Let us define the initial value of vapour supersaturation $\zeta _{0}$ by means of
\begin{equation}
\label{3.7}
\zeta _{0} \equiv \frac{n_{0} -n_{\infty } \left(T_{0} \right)}{n_{\infty } \left(T_{0} \right)} .
\end{equation}
Let us introduce an important dimensionless parameter $k$ in accordance with
\begin{equation}
\label{3.8}
k \equiv \left(\frac{q}{k_{B} T_{0} } \right)^{2} \frac{k_{B} Dn_{0} }{\kappa } .
\end{equation}
Taking Eqs. \eqref{2.13} and \eqref{3.5a} into consideration, from Eq. \eqref{3.8} we can obtain the following estimate:
\begin{equation}
\label{3.9}
k \sim \left(\frac{q}{k_{B} T_{0} } \right)^{2} \frac{n_{0} }{n_{g} c_{g} }.
\end{equation}
From Eqs. \eqref{3.8} and \eqref{3.9} it can be seen that parameter $k$ depends on the concentration of the inert gas $n_{g} $ in inverse proportion to $n_{g} $. Beside the dimensionless condensation heat $q/k_{B} T_{0}$, values $\zeta _{0} $ and $k$ can be also seen as initial (prior to the nucleation of the droplet) parameters of the vapour-gas medium.

If we combine Eq. \eqref{3.5} with the Clapeyron-Clausius formula Eq. \eqref{3.6}, and take into account definitions Eqs. \eqref{3.7} and \eqref{3.8}, for $T_{d} -T_{0} $ we will have a nonlinear transcendental equation
\begin{equation}
\label{4.3}
1 - \frac{q}{k_{B} T_{0} } \left(\frac{T_{d} -T_{0} }{T_{0} } \right)\frac{1}{k} =\frac{1}{1+\zeta _{0} } \exp \left(\frac{q}{k_{B} T_{0} } \frac{T_{d} -T_{0} }{T_{0} } \right).
\end{equation}
This equation is identical to Eq. (29) in \cite{Kuni1985}.

\section{Strong display of condensation heat release effects}
\label{Strong}

Far from the liquid-vapour critical point the strong inequality $q/k_{B} T_{0} \gg 1$ is usually observed. For this very reason, when the value of the ratio $n_{g}/n_{0}$ (see condition Eq. \eqref{1.3}) in not too large in comparison to 1, then the estimate Eq. \eqref{3.9} shows that
\begin{equation}
\label{4.1}
k \gg 1.
\end{equation}
Inequality Eq. \eqref{4.1}, as will be seen further from Eq. \eqref{4.7}, corresponds to a strong display of condensation heat release effects. From Eq. \eqref{3.9} it follows that in order for Eqs. \eqref{1.3} and \eqref{4.1} to be compatible there is a need for
\begin{equation}
\label{4.2}
1 \ll \frac{n_{g} }{n_{0} } \ll \frac{1}{c_{g} } \left(\frac{q}{k_{B} T_{0} } \right)^{2}.
\end{equation}
The stronger the inequality $q/k_{B} T_{0} \gg 1$, the weaker the upper bound on $n_{g}/n_{0}$ in Eq. \eqref{4.2}. Thus, in this section we will study the case when Eq. \eqref{4.1} is observed. The opposite case, where $k \ll 1$, will be studied in the next section. Let us stress that a formal transition towards the case with $k \ll 1$ in the formulas obtained from $k \gg 1$ is not justified.

Solving Eq. \eqref{4.3} at the observation of the inequality Eq. \eqref{4.1} by means of the method suggested in \cite{Kuni1985}, which does not imply the smallness of exponent in Eq. \eqref{3.6}, we have
\begin{equation}
\label{4.4}
\frac{T_{d} -T_{0} }{T_{0} } = \left(\frac{k_{B} T_{0} }{q} \right)\frac{k}{k+1} \ln \left(1+\zeta _{0} \right) ~~~~~ \left(k\gg 1\right).
\end{equation}
It is easy to make sure that Eq. \eqref{4.4} is the solution to Eq. \eqref{4.3} at the observation of the inequality Eq. \eqref{4.1} and at $\ln \left(1+\zeta _{0} \right)\sim 1$ (for the initial vapour supersaturation $\zeta_0$ it is usually correct just that $\zeta_0 \simeq 3 \div 5$: then the droplet can nucleate fluctuationally).

From Eq. \eqref{4.4}, taking $q/k_{B} T_{0} \gg 1$ and $\ln \left(1+\zeta _{0} \right)\sim 1$ into account, the inequality $(T_{d} -T_{0}) /T_{0} \ll 1$ (used in Eq. \eqref{3.6}) follows. From Eq. \eqref{4.4} there is also another conclusion, which might seem unexpected at the first sight. We will deal with that one further, after Eq. \eqref{5.1}. However, the stronger the inequality $q/k_{B} T_{0} \gg 1$ , the less is $(T_{d} -T_{0}) /T_{0}$ in accordance with Eq. \eqref{4.4}. Analytical expression Eq. \eqref{4.4} is an identical equivalent to a similar expression Eq. (45) in \cite{Kuni1985}

From Eqs. \eqref{3.5} and \eqref{4.4}, taking Eqs. \eqref{3.7} and \eqref{3.8} into account, it identically follows that
\begin{eqnarray}
\label{4.5}
n_{0} -n_{\infty } \left(T_{d} \right)=\frac{\left(1+\zeta _{0} \right)\ln \left(1+\zeta _{0} \right)}{\left(k+1\right)\zeta _{0} } \left[n_{0} -n_{\infty } \left(T_{0} \right)\right] \nonumber \\
\left(k \gg 1\right).
\end{eqnarray}
This particularly shows that $n_{0} -n_{\infty } \left(T_{d} \right)>0$ at $n_{0} -n_{\infty } \left(T_{0} \right)>0$, despite $n_{\infty } \left(T_{d} \right)>n_{\infty } \left(T_{0} \right)$ at $T_{d} >T_{0} $.

On the basis of Eqs. \eqref{1.17} and \eqref{3.4} we have equalities
\begin{equation}
\label{4.6}
\frac{dR^{2} / dt}{\left(dR^{2}/dt \right)^{\left(0\right)} } = \frac{b}{b^{\left(0\right)} } = \frac{a}{a ^{\left(0\right)} },
\end{equation}
where the upper index $\left(0\right)$ marks values in the absence of condensation heat release effects at the same diffusion coefficient $D$. The ratio in the left part of Eq. \eqref{4.6} characterizes the degree of change in droplet growth rate by means of heat release condensation effects. Developing the right part of Eq. \eqref{4.6} by means of Eqs. \eqref{1.27} and \eqref{3.1}, and making use of Eq. \eqref{4.5}, we have
\begin{equation}
\label{4.7}
\frac{dR^{2}/dt} {\left(dR^{2}/dt \right)^{\left(0\right)} } =\frac{\left(1 + \zeta_{0} \right)\ln \left(1+\zeta _{0} \right)}{\left(k+1\right)\zeta_{0} } ~~~~~~~~~~ \left(k \gg 1\right).
\end{equation}
Analytical expression Eq. \eqref{4.7} is an identical equivalent to a similar expression Eq. (51) in \cite{Kuni1985}. The stronger the inequality Eq. \eqref{4.1}, the more significant, according to Eq. \eqref{4.7}, is the decline in the rate of droplet growth as a result of heat release condensation effects (though simultaneously from Eq. \eqref{4.4} it follows that value $(T_{d} -T_{0})/T_{0}$ at given $q/k_{B} T_{0}$ will be practically constant). Inequality Eq. \eqref{4.1} is consequently responsible for the strong display of condensation heat release effects.

At the same time we are arriving at an important conclusion. The constant in time temperature of the growing droplet, the only condition for the existence of self-similar solutions Eqs. \eqref{1.24} and \eqref{2.15}, coincides with the constant temperature of the droplet that settles, in accordance with \cite{Kuni1985}, simultaneously with the settlement of the diffusion regime of the droplet. Constant temperature of the growing droplet, which is required by the self-similar theory, is consequently prominent from a physical standpoint, or stable, to be more exact.

By using Eqs. \eqref{1.27}, \eqref{3.4}, \eqref{3.7} and \eqref{4.5}, we can reduce the self-similar solution from Eq. \eqref{1.24} to
\begin{eqnarray}
\label{4.8}
n\left(\rho \right)=n_{0} \left[1-\frac{\ln \left(1+\zeta _{0} \right)}{k+1} \int _{\rho }^{\infty }\frac{dx}{x^{2} } \exp \left(-\frac{a}{2} x^{2} \right) \right] \nonumber \\ 
\left(k\gg 1\right),
\end{eqnarray}
and the self-similar solution from Eq. \eqref{2.15} to
\begin{eqnarray}
\label{4.9}
T\left(\rho \right)=T_{0} \left[1+\frac{qDn_{0} }{\kappa T_{0} } ~\frac{\ln \left(1+\zeta _{0} \right)}{k+1} 
\int _{\rho }^{\infty }\frac{dx}{x^{2} } \exp \left(-\frac{a D}{2 \chi} x^{2} \right) \right] \nonumber \\  
~~~~~~~~~ \left(k\gg 1\right).
\end{eqnarray}
By reason of Eqs. \eqref{1.12}, \eqref{1.13} and \eqref{1.16}, Eq. \eqref{4.8} complies with the diffusion equation Eq. \eqref{1.4} and with all the boundary conditions Eqs. \eqref{1.6}, \eqref{1.7} and \eqref{1.25}.

With provision for definition Eq. \eqref{3.8}, Eq. \eqref{4.9} may be rewritten in the form of
\begin{eqnarray}
\label{4.12}
T\left(\rho \right)=T_{0} \left[1+\frac{k_{B} T_{0} }{q} \frac{k}{k+1} \ln \left(1+\zeta _{0} \right)\int _{\rho }^{\infty }\frac{dx}{x^{2} } \exp \left(-\frac{a D}{2 \chi} x^{2} \right) \right] \nonumber \\
\left(k\gg 1\right).
\end{eqnarray}
By reason of Eqs. \eqref{1.12}, \eqref{1.16}, \eqref{2.5}, \eqref{3.8} and \eqref{4.4}, Eq. \eqref{4.9} complies with the heat conductivity equation Eq. \eqref{2.1} and with boundary conditions Eqs. \eqref{2.4}, \eqref{2.7} and \eqref{2.8}.

From Eqs. \eqref{1.27}, \eqref{3.1}, \eqref{3.4} and \eqref{4.5}, with due account for definition Eq. \eqref{3.7}, for value $a$ in Eqs. \eqref{4.8}  -- \eqref{4.12} and parameter $b$ in law in Eq. \eqref{1.17} on the growth of the droplet radius in time, we have:
\begin{equation}
\label{4.10}
a = b = \frac{n_{0} }{n_{l} } \frac{\ln \left(1+\zeta _{0} \right)}{k+1} ~~~~~~~~ \left(k\gg 1\right).
\end{equation}
Eqs. \eqref{4.8} -- \eqref{4.12} conjointly with Eq. \eqref{4.10} express self-similar solutions $n\left(\rho \right)$ and $T\left(\rho \right)$ directly via initial (prior to the nucleation of the droplet) parameters of the vapour-gas medium. Eqs. \eqref{4.8} -- \eqref{4.12}, crucial for the problem of the nearest-neighbour droplet \cite{Djikaev, Reiss}, previously have not been obtained.

Formula Eq. \eqref{4.10} in accordance with Eq. \eqref{1.21} makes it possible to obtain the following expression for time $t_{D}$ that passes after the nucleation of the droplet till the settlement of the diffusion regime of the droplet growth, at which the squared radius of the droplet becomes proportional to time:
\begin{equation}
\label{4.13}
t_{D} \simeq R_{D} ^{2} \left[\frac{n_{l} }{n_{0} } \frac{k+1}{2D\ln \left(1+\zeta _{0} \right)} \right] ~~~~~~~~ \left(k\gg 1\right).
\end{equation}
Eq. \eqref{4.13} was not previously obtained.

\section{Weak display of condensation heat release effects}
\label{Weak}

For the sake of completeness let us also consider the case of weak display of condensation heat release effects. In this case, instead of inequality Eq. \eqref{4.1} the reverse inequality will be observed:
\begin{equation}
\label{5.0}
k \ll 1.
\end{equation}
Solving Eq. \eqref{4.3}, that is observed irrespectively to the value of parameter $k$, by means of a petrurbation method at $k \ll 1$, we have
\begin{equation}
\label{5.1}
\frac{T_{d} -T_{0} }{T_{0} } =\left(\frac{k_{B} T_{0} }{q} \right)\frac{k\zeta _{0} }{1+\zeta _{0} } \left(1-\frac{k}{1+\zeta _{0} } \right) ~~~~~~~ \left(k\ll 1\right),
\end{equation}
which can be also confirmed by a direct verification. From Eq. \eqref{5.1} and $q/k_{B} T_{0} \gg 1$ it may be seen that at $k\ll 1$ the inequality $\left(T_{d} -T_{0} \right) /T_{0} \ll 1$ turns out to be a very strong one.

Let us notice that due to Eq. \eqref{3.8} value $\left(T_{d} -T_{0} \right) /T_{0}$ in Eq. \eqref{5.1} increases with the growth of  dimensionless condensation heat $q/k_{B} T_{0}$, while in Eq. \eqref{4.4} value $\left(T_{d} -T_{0} \right) /T_{0}$, on the contrary, declines with the growth of $q/k_{B} T_{0} \gg 1$. This difference is attributed to the nonlinearity of Eq. \eqref{4.3}.

Replacing Eq. \eqref{4.4} with Eq. \eqref{5.1}, instead of Eq. \eqref{4.5} and Eq. \eqref{4.7}, that are valid at $k\gg 1$, we have
\begin{equation}
\label{5.2}
n_{0} -n_{\infty } \left(T_{d} \right)=\left(1-\frac{k}{1+\zeta _{0} } \right)\left[n_{0} -n_{\infty } \left(T_{0} \right)\right] ~~~~~~~ \left(k\ll 1\right),
\end{equation}
\begin{equation}
\label{5.3}
\frac{dR^{2} / dt}{\left(dR^{2} /dt \right)^{\left(0\right)} } =1-\frac{k}{1+\zeta _{0} } ~~~~~~~~ \left(k \ll 1\right),  \end{equation}
that are valid at $k \ll 1$. In accordance with Eqs. \eqref{5.2} and \eqref{5.3}, heat release condensation effects are displayed weakly at $k \ll 1$.

Similarly, replacing Eq. \eqref{4.4} with Eq. \eqref{5.1}, instead of Eqs. \eqref{4.8} -- \eqref{4.13}, that are valid at $k\gg 1$, we have
\begin{widetext}
\begin{equation}
\label{5.4}
n\left(\rho \right)=n_{0} \left[1-\frac{\zeta _{0} }{1+\zeta _{0} } \left(1-\frac{k}{1+\zeta _{0} } \right)\int _{\rho }^{\infty }\frac{dx}{x^{2} } \exp \left(- \frac{a}{2} x^{2} \right) \right] ~~~~~~~ \left(k\ll 1\right),
\end{equation}
\begin{equation}
\label{5.5}
T\left(\rho \right)=T_{0} \left[1+\frac{qDn_{0} }{\kappa T_{0} } \frac{\zeta _{0} }{1+\zeta _{0} } \left(1-\frac{k}{1+\zeta _{0} } \right)\int _{\rho }^{\infty }\frac{dx}{x^{2} } \exp \left(-\frac{a D}{2 \chi} x^{2} \right) \right] ~~~~~~ \left(k\ll 1\right),
\end{equation}
\begin{equation}
\label{5.9}
T\left(\rho \right)=T_{0} \left[1+\frac{k_{B} T_{0} }{q} \frac{k\zeta _{0} }{1+\zeta _{0} } \left(1-\frac{k}{1+\zeta _{0} } \right)\int _{\rho }^{\infty }\frac{dx}{x^{2} } \exp \left(-\frac{a D}{2 \chi} x^{2} \right) \right]  ~~~~~~~~ \left(k\ll 1\right),
\end{equation}
\begin{equation}
\label{5.6}
a = b = \frac{n_{0} }{n_{l} } \frac{\zeta _{0} }{1+\zeta _{0} } \left(1-\frac{k}{1+\zeta _{0} } \right) ~~~~~~~~ \left(k\ll 1\right),
\end{equation}
\begin{equation}
\label{5.8}
t_{D} \simeq R_{D} ^{2} \left[\frac{n_{l} }{n_{0} } ~\frac{1+\zeta _{0} }{2D\zeta _{0} } \left(1+\frac{k}{1+\zeta _{0} } \right)\right] ~~~~~~~~~~~ \left(k\ll 1\right),
\end{equation}
\end{widetext}
that are valid at $k \ll 1$.
Eqs. \eqref{5.1} -- \eqref{5.8} include only initial (prior to the nucleation of the droplet) parameters of the vapour-gas medium. Eqs. \eqref{5.4} -- \eqref{5.9}, crucial for the problem of the nearest-neighbour droplet \cite{Djikaev, Reiss}, previously have not been obtained, neither in \cite{Kuni1985}. Within limits of $k = 0$, Eqs. \eqref{5.4}, \eqref{5.6} and \eqref{5.8} correspond to isothermal vapour condensation at the absence of heat effects.

It is easy to make sure that Eq. \eqref{5.4} by reason of Eq. \eqref{5.2} and definition Eq. \eqref{3.7} complies with the boundary condition Eq. \eqref{1.23} on the surface of the droplet. It is obvious that Eq. \eqref{5.4} also complies with the boundary condition Eq. \eqref{1.25} in the infinite distance from the droplet. For this very reason $n\left(\rho \right)$ changing in Eq. \eqref{5.4} within the interval of $n_{\infty } \left(T_{d} \right)\le n\left(\rho \right)\le n_{0} $, which has the width of $n_{0} -n_{\infty } \left(T_{d} \right)$ set by the Eq. \eqref{5.2}. This width exceeds significantly the one that is provided in Eq. \eqref{4.5}, i. e. width $n_{0} -n_{\infty } \left(T_{d} \right)$ of the interval where $n\left(\rho \right)$ changing in Eq. \eqref{4.8} in case of $k\gg 1$.

It is also easy to make sure that Eq. \eqref{5.9} by reason of Eq. \eqref{5.1} complies with the boundary condition Eq. \eqref{2.7} on the surface of the droplet. It is obvious that Eq. \eqref{5.9} also complies with the boundary condition Eq. \eqref{2.8} in the infinite distance from the droplet. For this very reason $T \left(\rho \right)$ changing in Eq. \eqref{5.9} within the interval of $T_{0} \le T\left(\rho \right)\le T_{d} $, which has the width of $T_{d} -T_{0} $ set by the Eq. \eqref{5.1}. This width is much smaller than the one that is provided in Eq. \eqref{4.4}, i. e. width $T_{d} -T_{0} $ of the interval where $T\left(\rho \right)$ changing in Eq. \eqref{4.12} in case of $k\gg 1$.

Let us compare Eq. \eqref{4.8} valid at $k\gg 1$ with Eq. \eqref{5.4} valid at $k\ll 1$. The multiplier before the integral in Eq. \eqref{4.8} is smaller than multiplier before the integral in Eq. \eqref{5.4}. From Eq. \eqref{4.10} and Eq. \eqref{5.6} it is obvious that value $a$ in Eq. \eqref{4.8} will be also smaller than value $a$ in Eq. \eqref{5.4}, and, which is important, it will be smaller by an equal factor. At this, the integral in Eq. \eqref{4.8} will, on the contrary, exceed the integral in Eq. \eqref{5.4}. Vapour concentration $n\left(\rho \right)$ in Eq. \eqref{4.8} in case of $k\gg 1$, i.e. strong heat effects, will be denoted as  $n^{\left(1\right)} \left(\rho \right)$, and vapour concentration $n\left(\rho \right)$ in Eq. \eqref{5.4} in case of $k=0$, i.e. complete absence of heat effects, which will be considered further for the sake of simplicity, will be denoted by means of $n^{\left(2\right)} \left(\rho \right)$. From Eqs. \eqref{4.5}, \eqref{5.2}, by virtue of Eq. \eqref{3.7} and the boundary condition Eq. \eqref{1.23} on the surface of the droplet, we then have 
\begin{eqnarray}
\label{5.10}
n_{0} -\left. n^{\left(1\right)} \left(\rho \right)\right|_{\rho =1} =n_{0} \frac{\ln \left(1+\zeta _{0} \right)}{k+1} ~~~~ \left(k\gg 1\right), \nonumber \\ 
n_{0} -\left. n^{\left(2\right)} \left(\rho \right)\right|_{\rho =1} =n_{0} \frac{\zeta _{0} }{1+\zeta _{0} } ~~~~~~ \left(k=0\right).
\end{eqnarray}
On the other hand, at $\rho =\infty $ directly from Eqs. \eqref{4.8} and \eqref{5.4} we have
\begin{eqnarray}
\label{5.11}
\left. n^{\left(1\right)} \left(\rho \right)\right|_{\rho =\infty } =n_{0} ~~~~~ \left(k\gg 1\right), \nonumber \\ 
 \left. n^{\left(2\right)} \left(\rho \right)\right|_{\rho =\infty } =n_{0} ~~~~~ \left(k=0\right).
\end{eqnarray}

\begin{figure}
\includegraphics[width=60mm]{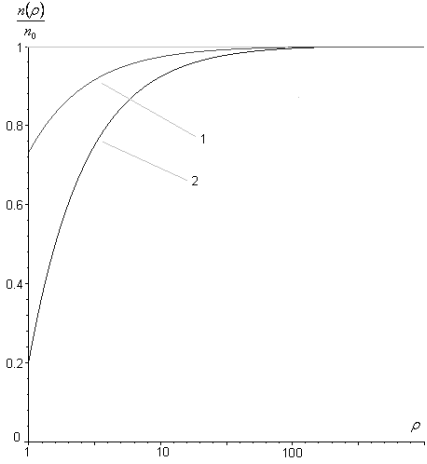}
\caption{The dependency of dimensionless vapour concentrations $n^{(1)} (\rho) / n_0$ (Curve 1) and $n^{(2)} (\rho) / n_0$ (Curve 2) on dimensionless distance $\rho $ at $n_{0}/2n_{l} = 10^{-5} $}
\label{Fig1}
\end{figure}

\begin{figure}
\includegraphics[width=60mm]{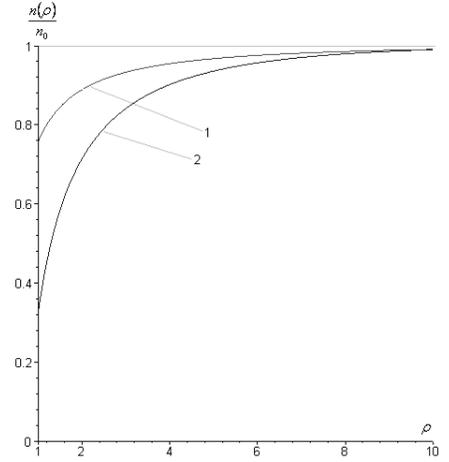}
\caption{ The dependency of dimensionless vapour concentrations $n^{(1)} (\rho) / n_0$ (Curve 1) and $n^{(2)} (\rho) / n_0$ (Curve 2) on dimensionless distance $\rho $ at $n_{0}/2n_{l} = 10^{-2} $}
\label{Fig2}
\end{figure}

More precise data on vapour concentration $n^{\left(1\right)} \left(\rho \right)$ and $n^{\left(2\right)} \left(\rho \right)$ can be obtained by numerical evaluation. The results of calculations of Eqs. \eqref{4.8} and \eqref{5.4} that represent the correlation between dimensionless vapour concentrations $n^{(1)} (\rho)/n_{0}$, $n^{(2)} (\rho)/n_{0}$ and dimensionless distance $\rho $ to the centre of the droplet in the units of droplet radius are represented by curves 1 and 2 in Figures \ref{Fig1} and \ref{Fig2}.

Values $a$ in Eqs. \eqref{4.8} and \eqref{5.4} were accordingly obtained by means of Eqs. \eqref{4.10} and \eqref{5.6}. In both Figure \ref{Fig1} and Figure \ref{Fig2} in Eqs. \eqref{4.8} and \eqref{4.10} it is assumed that $k=5$ (in the case of strong heat effects represented by curve 1), while in Eqs. \eqref{5.4} and \eqref{5.6} it is assumed that $k=0$ (in the case of a complete absence of heat effects represented by curve 2). In both Figure \ref{Fig1} and Figure \ref{Fig2} it is accepted that $\zeta _{0} =4$ (the initial vapour supersaturation $\zeta _{0}$ is already so high that a droplet can only nucleate in a fluctualtional way). In Figure \ref{Fig1} it is asserted that $n_{0}/2n_{l} =10^{-5}$, which corresponds to a significant remoteness from the critical liquid-gas point, while in Figure \ref{Fig2} it is asserted that $n_{0}/2n_{l} =10^{-2}$, which corresponds to an insignificant remoteness from the critical liquid-gas point.

One can also witness the concordance between Figs. \ref{Fig1} and \ref{Fig2} and Eqs. \eqref{5.10} and \eqref{5.11}. Let us notice that within $1 \le \rho \lesssim 2^{1/2} / 4 a^{1/2} $ Eqs. \eqref{4.8} and \eqref{5.4} comply with the steady theory with high precision (within these limits the absolute value of the exponent in Eqs. \eqref{4.8} and \eqref{5.4} is smaller than $1/16$, and the exponents themselves are rather close to 1). Thus, it can be seen that $n^{(1)} (\rho)/n_{0}$ and $n^{(2)} (\rho)/n_{0}$  draw near 1 in Fig.\ref{Fig1} (where $n_{0}/2n_{l} =10^{-5}$) still within the limits of the steady theory, which is correct at all values of  $\rho $ given in the logarithmic scale in Fig.\ref{Fig1}. In Fig.\ref{Fig2} (where $n_{0}/2n_{l} =10^{-2}$) the above mentioned curves draw near 1 only within the limits of the nonsteady theory, which is correct already when $\rho \gtrsim 3$.

It is appropriate to make the following comment: in order to find $dR^2(t)/dt$, as can be seen from Eqs. \eqref{1.17} and \eqref{1.18}, it is enough to know $n\left(\rho \right)$ only in the infinitely narrow vicinity of the droplet. However, in the problem of the nearest neighbour droplet \cite{Djikaev, Reiss} one also needs to know $n\left(\rho \right)$ in the distance from the droplet. In the case when $n_{0}/2n_{l} =10^{-2}$ (see Fig.\ref{Fig2}), the profile $n\left(\rho \right)$ will be steady only when $\rho \lesssim 3$. Though exactly then the steady approximation will be sufficient to obtain $dR^2(t)/dt$, it wont be sufficient in the problem of the nearest neighbour droplet. Consequently a nonsteady theory is needed to obtain self-similar Eqs. \eqref{4.8}, \eqref{4.12}, \eqref{5.4} and \eqref{5.9}. In Fig.\ref{Fig2} the nonsteady theory reveals itself already when $\rho \gtrsim 3$.

\begin{figure}
\includegraphics[width=60mm]{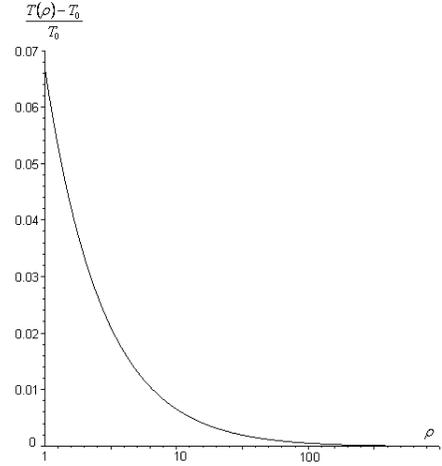}
\caption{ The dependency of the relative change of temperature of the vapour-gas mixture $(T(\rho)-T_{0})/T_{0}$ on dimensionless distance $\rho$ at $n_{0}/2n_{l} = 10^{-5}$}
\label{Fig3}
\end{figure}

\begin{figure}
\includegraphics[width=60mm]{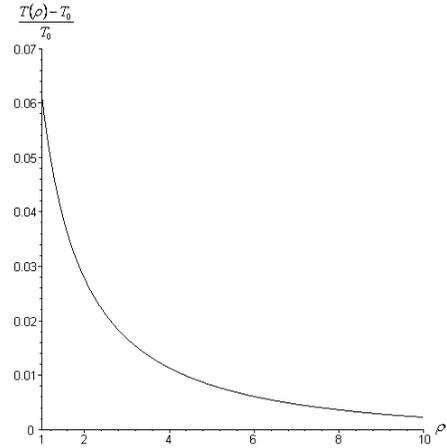}
\caption{The dependency of the relative change of temperature of the vapour-gas mixture $(T(\rho)-T_{0})/T_{0}$ on dimensionless distance $\rho$ at $n_{0}/2n_{l} = 10^{-2}$}
\label{Fig4}
\end{figure}

In addition, in Fig.\ref{Fig3} and \ref{Fig4}, when studying the case of strong heat effects, we have presented a plot of relative deviation of temperature of the vapour-gas medium $\left[T(\rho)-T_{0} \right]/T_{0}$ versus dimensionless distance $\rho $ to the centre of the droplet in the units of the radius of the droplet. Value $T\left(\rho \right)$ is obtained using Eq. \eqref{4.12}; and value $a$ -- by means of Eq. \eqref{4.10}. In the calculations it is assumed that $q/k_{B} T_{0} =20$, $k=5$, $\zeta _{0} =4$ and $D / \chi =1$. In Fig.\ref{Fig3} it is accepted that $n_0/2n_l =10^{-5} $; and in Fig.\ref{Fig4} it is accepted that $n_0/2n_l =10^{-2} $ In Fig.\ref{Fig3} the values of $\rho $ are given on a logarithmic scale.

The fact that the droplet heats the surrounding medium, which primarily consists of the inert gas, results in a dependence of diffusion coefficient $D$ of vapour molecules in the inert gas on the medium temperature $T$. Using the gas-kinetic equation $D \simeq \lambda v_T / 3$, where $v_T $ is the mean thermal velocity of a vapour molecule, considering the inert gas to be ideal and remaining at the given pressure, we can easily make sure that the coefficient $D$ depends on $T$ in direct proportion to $T^{3/2}$. If we denote the coefficient of diffusion, which depends on $T$, by means of $D_{T} $, we will then have $D_{T} =(T/T_0)^{3/2} D$, where $D$ is the previously known coefficient of diffusion at temperature $T_{0} $ in the distance from the droplet. Supposing that $(T - T_0)/T_0 \ll 1$, for the relative increase of coefficient $D_{T} $ with the increase of temperature $T$ on drawing near the droplet we consequently have
\begin{equation}
\label{5.12}
(D_{T} -D)/D =3(T-T_{0})/2T_{0} .
\end{equation}

As Fig. \ref{Fig3} and \ref{Fig4} show, even at the strong display of condensation heat release effects and even on the very surface of the droplet, when $(T-T_{0}) / T_{0}$ is the highest, the following will still be correct: ${(T-T_{0}) / T_{0} } < 0.07$. This, in accordance with Eq. \eqref{5.12}, provides for $(D_{T} -D)/D < 0.1$. Then, the measure of inaccuracy brought in the considered theory will be small and wont exceed 10\%. Let us also note that the measure of inaccuracy concerned with the linearization by $(T-T_{0}) / T_{0}$  of the exponent in the Clapeyron-Clausius formula Eq. \eqref{3.6} is of the same order. For this very reason the dependance of the diffusion coefficient on temperature can be neglected (it is referred to temperature $T_0$).

\section{Estimates of common characteristics of a nonisothermal vapour condensation on a droplet growing in diffusion regime}
\label{Estimates}

Let us estimate common characteristics of a nonisothermal vapour condensation for the case of strong and for the case of weak display of condensation heat release effects by the example of a particular system  water vapour in the air.

In accordance with Eqs. \eqref{1.1}, \eqref{1.21} and \eqref{1.17} let us write
\begin{equation}
\label{6.1}
R_D \simeq 10 \lambda, ~~~t_D \simeq R_D^2/(dR^2(t)/dt),
\end{equation}
where it is assumed that $N \simeq 10$. Let us accept that $T_0 = 273~K$ and $\zeta_0 = 4$. The following expressions are of no dependence on pressure values $p_g$ of the inert gas:
\begin{eqnarray}
\label{6.2}
n_{\infty}(T_0) = 1.6 \cdot 10^{17} ~cm^{-3}, ~~~ n_0 = 8 \cdot 10^{17} ~cm^{-3}, \nonumber \\
n_l = 3.3 \cdot 10^{22} ~cm^{-3}, ~~~~ \kappa = 2.62 \cdot 10^3 ~\frac{erg}{cm \cdot s \cdot K}, \nonumber \\
\frac{q}{k_B T_0} = ~17.9.
\end{eqnarray}
Here the data on $n_{\infty}(T_0)$, $\kappa$, $q/k_B T_0$ is taken from \cite{Handbook} and equation $\zeta_0 = 4$ and the definition from Eq. \eqref{3.7} are accounted for.

In the case of strong display of condensation heat release effects we shall assume that $p_g = 0.5 ~atm \simeq 5 \cdot 10^5 ~dyne/cm^2$. By means of constitutive equation of the ideal gas with $n_0$ obtained above, we have $n_g/n_0 = 17$; thus the condition Eq. \eqref{1.3} is very well observed. In accordance with \cite{Handbook}, for $D$ at $p_g = 1 ~atm$ the following is correct: $D = 2.05 \cdot 10^{-1} ~cm^2/s$. As $D$ depends on $p_g$ in inverse proportion, then at $p_g = 0.5 ~atm$ we will have $D = 4.1 \cdot 10^{-1} ~cm^2/s$. Then from Eq. \eqref{3.8}, with regard to Eq. \eqref{6.2}, we have $k = 5.6$, and then from Eq. \eqref{4.10}, with regard to Eq. $\zeta_0 = 4$ and Eq. \eqref{6.2}, we have $a = b = 5.9 \cdot 10^{-6}$. Simultaneously from Eq. \eqref{1.17} we have $dR^2(t)/dt \simeq 4.8 \cdot 10^{-6} ~cm^2/s$. Value $k = 5.6$ complies perfectly with Eq. \eqref{4.1}, i. e. the condition of strong display of heat effect. Further, when we have found $D$ from the gas-kinetic formula $D \simeq \lambda v_T / 3$ taking into account an easily obtained value $v_T = 5.7 \cdot 10^4 ~cm/s$ of the heat velocity of a water vapour molecule, we can come to $\lambda \simeq 2.2 \cdot 10^{-5} ~cm$. Then, from Eq. \eqref{6.1}, when the value $dR^2/dt$ has been already obtained, we have $R_D \simeq 2.2 \cdot 10^{-4} ~cm$ and $t_D \simeq 1.0 \cdot 10^{-2} ~s$.

In the case of weak display of condensation heat release effects we will still assume that $T_0 = 273 ~K$ and $\zeta_0 = 4$. However, we will suppose that $p_g = 10 ~atm \simeq 10^7 ~dyne/cm^2$, i. e. we will increase the pressure of the inert gas 20-fold. Despite this fact, the data from Eq. \eqref{6.2} will still be valid due to its  rather low sensitivity to $p_g$ (at $n_l \gg n_0$).  Now, due to the constitutive equation of the ideal gas we will have $n_g/n_0 = 3.4 \cdot 10^2$, thus the condition Eq. \eqref{1.3} is very well observed. Simultaneously we will have $D = 2.05 \cdot 10^{-2} ~cm^2/s$, and, considering Eqs. \eqref{3.8}, \eqref{5.6} and \eqref{1.17} with regard to \eqref{6.2}, we will also have $k = 0.28$, $a = b = 1.9 \cdot 10^{-5}$ and $dR^2(t)/dt = 7.8 \cdot 10^{-7}$. Value $k = 0.28$ complies with the condition of weak display of heat effect (at $k = 0.28$ and $\zeta_0 = 4$ one can neglect a correctional member in terms of $k$ in Eq. \eqref{5.6}). Finally, at $D = 2.05 \cdot 10^{-2} ~cm^2/s$, and $v_T = 5.7 \cdot 10^4 ~cm/s$ using the gas-kinetic formula $D \simeq \lambda v_T / 3$ we have $\lambda \simeq 1.1 \cdot 10^{-6} ~cm$, and then by means of relations in Eq. \eqref{6.1} we have $R_D \simeq 1.1 \cdot 10^{-5} ~cm$ and $t_D \simeq 1.6 \cdot 10^{-4} ~s$.

At the transition from the case of strong heat effect to the case of weak heat effect the derivative $dR^2(t)/dt$, notwithstanding Eq. \eqref{4.7}, is decreasing. This fact can be explained as now during this transition we see a significant decrease of the coefficient of diffusion $D$, which was considered constant in Eq. \eqref{4.7}.

\begin{acknowledgments}
The research has been carried out with the financial support of the Russian Analytical Program ``The Development of Scientiﬁc Potential of Higher Education'' (2006 -- 2008), project RNP.2.1.1.1812. Fundamental Problems of Physics and Chemistry of
Ultradisperse Systems and Interfaces.
\end{acknowledgments}

\newpage
\bibliography{Manuscript}

\end{document}